\documentclass[twocolumn]{aa}
\usepackage{graphicx}
\usepackage{natbib}
\newcommand{\lsim}{\,\raisebox{0.2em}{$<$}\!\!\!\!\!
\raisebox{-0.25em}{$\sim$}\,}
\newcommand{\gr}{$\gamma$-ray \,}
\newcommand{\grs}{$\gamma$-rays \,}
\newcommand{\rxj}{RX~J1713.7-3946 \,}
\newcommand{\hess}{H.E.S.S.}
\begin{document}

   \title{Hadronic versus leptonic origin of the gamma-ray emission
   from Supernova Remnant RX~J1713.7-3946 \,}

%   \subtitle{}

\authorrunning{Berezhko \& V\"olk}  
\titlerunning{Hadronic vs leptonic \grs in SNR \rxj} 

\author{E.G. Berezhko \inst{1}
          \and H.J. V\"olk \inst{2}
}

  \institute{Yu.G. Shafer Institute of Cosmophysical Research and Aeronomy,
                     31 Lenin Ave., 677980 Yakutsk, Russia\\
              \email{berezhko@ikfia.ysn.ru}
         \and
           Max Planck Institut f\"ur Kernphysik,
                Postfach 103980, D-69029 Heidelberg, Germany\\
              \email{Heinrich.Voelk@mpi-hd.mpg.de}           
}

   \offprints{H.J. V\"olk}

   \date{Received month day, year; accepted month day, year}

   \abstract{} 
{The hadronic vs. leptonic origin of the gamma-ray
emission from the Supernova Remnant RX J1713.7-3946 is discussed both
in the light of new observations and from a theoretical point of
view.}  
{The existing good spatial correlation of the gamma-ray emission and
the nonthermal X-ray emission is analyzed theoretically. In addition,
the recently published new H.E.S.S. observations define the energy
spectrum more precisely, in particular at the high and low energy
ends of the instrument's dynamical range. There now exist
much more constraining X-ray observations from Suzaku that extend
substantially beyond 10 keV. These new data are compared with the
authors' previous theoretical predictions, both for dominant hadronic
and for simple inverse Compton models.}
{Apart from the well-known MHD correlation between magnetic field
strength and plasma density variations, emphasized by the
wind-bubble-structure of the remnant, it is argued that the
regions of magnetic field amplification also are correlated with enhanced
densities of accelerated nuclear particles and the associated
streaming instabilities. Therefore a correlation of nonthermal X-ray
and \gr emission is not only possible but even to be expected for a
hadronic emission scenario. A leptonic origin of the gamma-ray
emission would require an implausibly uniform strength of the magnetic
field. The observational and theoretical inferences about substantial
field amplification in this remnant agree very well with the recent
X-ray and \gr observations.}  
{All this argues strongly for the dominance of hadronic \grs in the
\gr emission spectrum and a fortiori for an overwhelming contribution
of nuclear cosmic ray particles to the nonthermal energy in this
remnant.}

  
\maketitle

%
%________________________________________________________________

\section{Introduction}
Hard X-ray emission from young Supernova Remnants (SNRs) was first detected in
SN 1006 and interpreted as synchrotron emission from electrons in the energy
range of tens of TeV \citep{koya95}. Such emission meanwhile has been found in
many young shell-type objects. These electrons are also expected to produce
very high energy (VHE) \grs in the $> 100$~GeV range by inverse Compton (IC)
collisions with low energy photons from the ambient radiation field, in
particular the Cosmic Microwave Background (CMB). In regions of high gas
density this may also be accompanied by a non-negligible emission of nonthermal
Bremsstrahlung \grs\ It is therefore not {\it a priori} clear whether the VHE
\grs detected in several shell-type SNRs up to now -- Cas~A
\citep{aha01a,al07}, \rxj \citep{mur00,enomoto02,aha04,aha06,aha07a}, RX
J0852.0-4622 \citep{Katagiri,aha05,aha07b,enomoto06}, and RCW~86 \citep{Hoppe}
-- are indeed the result of inelastic collisions of energetic nuclei with gas
atoms, as expected theoretically if a sizeable fraction of the SN explosion
energy goes into nuclear cosmic rays (CRs) \citep{drury94,bv97}, or whether
they are merely due to IC collisions of the X-ray synchrotron emitting
electrons.

The latter inference seems at least in a rough sense possible if the effective
magnetic field strength in these objects is sufficiently low. Then the energy
spectrum of accelerated electrons, calculated from the observed synchrotron
spectrum, contains sufficiently many particles so that a strong enough IC \gr
emission results. Since magnetic fields are difficult to determine, a possible
phenomenological approach is to assume that the interior (postshock) magnetic
field in the SNR is the result of MHD-compression of a circumstellar field
whose strength is at best a few $\mu$G. For \rxj and RX J0852.0-4622 such a
procedure works, at least as far as the overall {\it amplitude} of the \gr
spectrum is concerned. For Cas~A this is true as well. However, all these
remnants presumably correspond to core collapse SN explosions into the wind
bubble of a massive progenitor star, with a rather complex circumstellar
magnetic field structure. In contrast, for the remnants of several type Ia SNe
which could not be detected so far in $\gamma$-rays such as Tycho's SNR, SN
1006, and Kepler's SN, this scheme leads to large overpredictions for the \gr
flux, unless the mean interior magnetic field strength $B$ is assumed to be
significantly higher than the value $\sim 10~\mu$G expected from a $\lsim
5~\mu$G circumstellar field for a strong quasi-parallel, adiabatic shock with a
compression ratio of 4 (V\"olk et al. 2007, 2008). Since electron acceleration
cannot be responsible for such an amplification of the interstellar field, this
makes a pure electron acceleration scenario generally untenable in our view.

The acceleration of nuclear particles on the other hand cannot be deduced in a
similar way from some other electromagnetic emission -- even though the
magnetic field strength and the ion injection rate are inferred from the
{\it form} of the electron synchrotron spectrum \citep{bkv02}. The energetic
nuclear component needs to be calculated theoretically, and the amplitude of
the accelerated particle distribution can only result from a nonlinear
theory. This has been done successfully in a number of cases and shows that the
overall energy in accelerated nuclei is indeed a large fraction -- about $10$\%
-- of the available hydrodynamic explosion energy, and that the local
efficiency at those parts of the shock surface, where ion injection is
effective, amounts to even about 50 \% \citep[see e.g.][for a review]{ber05}.

Given this high acceleration efficiency, at reasonable thermal gas
densities the expected rate of hadronic collisions with gas atoms and
the resulting \gr production by $\pi^0$-decay turns out to be
consistent with the observations in the \gr\-detected objects analyzed
so far: Cas~A \citep{bpv03}, \rxj \citep{bv06} [hereafter referred to
as BV06], \citep{bv07}, RX J0852.0-4622 \citep{bpv08}. Also for
the other objects that we have analyzed theoretically -- Tycho's SNR
\citep{vbkr02,vbk05,vbk08}, SN 1006 \citep{bkv02,kbv05}, and Kepler's
SNR \citep{bkv06} -- and for which only upper limits exist so
far, the hadronic \gr emission is expected to dominate the IC
fraction, even though for the low-density object SN 1006 only by a
modest margin.

Within the errors, the theoretically derived magnetic field strengths also
agree with those deduced from the observations of filamentary X-ray synchrotron
features, often found at the outer rims of the SNRs
\citep[e.g.][]{vl03,long,bam03,bkv03,bv04a,vbk05,par06}.  The field strengths
typically are an order of magnitude greater than the compressed
circumstellar field, substantially reducing the IC
spectrum amplitude for a given synchrotron spectrum and thus tending to yield a
negligible leptonic contribution to the observed VHE \gr flux.

Although the dominant acceleration of nuclear particles is clearly favored by
what has been said up to now, leptonic scenarios have been studied in some
detail \citep[e.g.][]{Porter06,Katz08}. Especially for the experimentally
best-studied SNR \rxj the question is whether an IC spectrum, scaled in
amplitude from the synchrotron spectrum in the sense described above, is
compatible with the {\it form} of the observed \gr spectrum. A second question
regards the observed spatial correlations of the morphology in hard X-rays,
assumed to be the result of synchrotron radiation, and in $\gamma$-rays. Such a
correlation has been observed in \rxj and might at first sight be attributed to
a common leptonic population of energetic particles. It remains to be seen
whether such an inference holds upon deeper scrutiny.

Another, not quite resolved question concerns the absence of thermal X-ray
emission in the two SNRs \rxj and RX J0852.0-4622 that are spatially resolved
at TeV energies. Also the radio synchrotron emission is very weak in comparison
to the strong X-ray synchrotron emission in both sources. Since the earliest
observations and analyses \citep{sla99}, the explanation for this situation has
been the assumption that the SN explosion occurred into the very rarefied
stellar wind bubble of a massive progenitor star. The model of BV06 specifies
the bubble structure in detail. It demonstrates that the main gas heating and
particle acceleration occurs beyond the wind region and bubble, when the SNR
shock propagates into the swept-up, radiatively cooled shell of interstellar
matter that was generated by the forward shock driven into the ambient ISM by
the expansion of the wind bubble. Therefore the wind bubble has two main
effects: the SNR can reach a large size quickly and it is then rather quickly
decelerated in the dense shell. During propagation of the SNR shock wave
through the wind and bubble the gas density is very low, resulting in a very
low thermal X-ray emission of this material. The shock's late encounter of the
massive shell heats the nuclear particles, but leaves little time for the
subsequent relaxation in the postshock region of the SNR that heats the thermal
electrons by Coulomb collisions with the heavy ions. In this way a minimum of
thermal X-ray emission from the remnant is combined with a maximum of hard
X-ray synchrotron and hadronic \gr emission from the SNR. In addition, the
shock is nonlinearly modified. For a SNR propagating into a uniform medium with
a uniform magnetic field this implies that the larger part of the shock surface
corresponds to a quasi-perpendicular shock with a strongly reduced injection of
nuclear particles \citep{vbk03}. Suprathermal injection of ions is only
possible in the quasi-parallel shock regions. If the spatial scales of the
quasi-perpendicular regions are large enough, the cross-field diffusion of the
highest-energy particles from the neighboring quasi-parallel shock regions also
does not reach deeply into these quasi-perpendicular regions. In the
corresponding magnetic flux tubes no magnetic field amplification occurs either
and the shock remains essentially un-modified there. This means that in the
quasi-perpendicular regions the shock dissipation and therefore the gas heating
occurs in a locally unmodified shock with the overall shock speed, leading to a
rather high gas temperature. In the case where a radiatively cooling shell of a
wind bubble is the main obstacle for the SNR expansion, the situation may be
different. The MHD instabilities of such a shell probably break it into many
small regions with strongly varying field directions. Then the spatial scales
separating the quasi-perpendicular from the quasi-parallel shock regions may
become small enough that crossfield diffusion can smear out the
quasi-perpendicular regions and particle acceleration occurs practically
everywhere over the shock surface \citep{vbk07,voelk08}. In the extreme this
implies shock modification over the entire shock region and thus a reduced gas
heating due to the subshock dissipation only\footnote{The given \gr flux
requires in addition a reduction of the gas density which reduces the
thermal emission.}. Compared to an un-modified shock
of the same overall speed, the gas temperature is then diminished by a
factor $ \approx \sigma_\mathrm{s}^2 / \sigma^2  \approx 0.25$,  where
$\sigma$ and $\sigma_\mathrm{s}$ denote the overall shock compression ratio and
the subshock compression ratio, respectively. This
reduces the emission of soft thermal X-rays drastically.

Such a configuration is not easily analysed in detail with standard methods of
X-ray astronomy. However, this also leaves uncertainties in the evaluation of
the model's validity. Here, as in BV06, we proceed under the assumption that
the concrete wind bubble model, or its eventual improvement, is consistent with
the fact that no thermal emission has been found up to now.

The purpose of the present paper is a discussion of the above two
questions for \rxj, even though we expect analogous arguments to
hold for the other objects mentioned.  In section 2 we discuss
the correlations between the gas density and the magnetic field
strength that are possible, or are even to be expected, in the
circumstellar medium before the SN explosion. In an approximate way we
also show how variations in gas density lead to correlated variations
in nuclear CR pressure and in the hadronic \gr emission. However, we shall
argue that the magnetic field {\it direction} influences
particle injection into the shock acceleration process, so that
circumstellar density enhancements are only a necessary but not a
sufficient condition for enhancements of the \gr emission. Under most
circumstances electron synchrotron emission is then expected to
increase with hadronic \gr emission as well. We also briefly
discuss the alternative purely leptonic scenario. In section 3 we
compare our calculations for the hadronic \gr emission, the
synchrotron emission, and for the IC and Bremsstrahlung \gr emission
with the latest X-ray and \gr measurements for \rxj. The results are
finally compared with a purely leptonic scenario, calculated by
assuming ion injection and thus also magnetic field amplification to
be negligibly small. It turns out that a hadronic origin of the \gr
emission is consistent with all measurements and with our theory, whereas
the leptonic scenario runs into serious difficulties. Section 4
contains our conclusions.

Very recently an independent discussion of these questions by \citet{tua08} has
come to our attention. Their paper is more observation oriented and is partly
complementary to ours. It uses phenomenological estimates for the accelerated
particle spectra. However their conclusions are similar to ours.

\section{Expected morphological correlations}

In this section we investigate the spatial correlations that arise from
compressions or expansions of the thermal gas with a frozen-in magnetic field,
in a given radiation field. In addition we argue that variations of the nuclear
energetic particle density -- which are most likely the result of spatial
variations of the injection rate into the acceleration process -- lead to
positively correlated enhancements of the magnetic field strength as a result
of magnetic field amplification by CR streaming instabilities.

\subsection{Compressions/de-compressions of the thermal plasma}

Consider briefly quasi-static compressions / de-compressions of the gas density
by plasma motions across the magnetic field in the MHD limit. 

In this case the conservation of
magnetic flux compresses the field together with the thermal gas, increasing at
the same time the target for electrons to produce synchrotron emission as well
as the target for energetic nuclei in inelastic collisions with the gas atoms.
In both cases the densities of the X-ray synchrotron emitting electrons and
of the \gr emitting nuclei vary approximately in the same way,
because the electrons and nuclei concerned have roughly the same
energy. Therefore their spatial distributions are essentially the same.

In the 1-fluid MHD-approximation the mass density $\rho$ and the
magnetic field vector $\vec{B}$ are related to the mass velocity $\vec{u}$
through the conditions of conservation of mass and magnetic flux:
\begin{equation} {\partial \rho \over \partial t} + {\vec{u} \cdot \nabla \rho}
  \equiv {\mathrm{d} \rho \over \mathrm{d} t} = -\rho \nabla \cdot \vec{u}
\label{eq1}
\end{equation}
\begin{equation}
{\partial \vec{B} \over \partial t} = \nabla \times (\vec{u} \times \vec{B}).
\label{eq2}
\end{equation}

\noindent Being interested primarily in the variations of $\rho$ and
the magnitude $B$ of $\vec{B}$, it is more useful for our considerations to
reduce Eqs.(1-2) to
equations for $B^2$ and $B/\rho$ alone, using $\nabla \cdot \vec{B}=0$:
\begin{equation}
{{1 \over 2B^2} {\mathrm{d} B^2 \over \mathrm{d} t}}
= - \nabla \cdot \vec{u}- {\vec{B} \cdot (\vec{B} \cdot \nabla)\vec{u}}
\label{eq3}
\end{equation}
\begin{equation}
{\mathrm{d} ln{(B / \rho)} \over \mathrm{d} t}
= B^{-2} \vec{B} \cdot (\vec{B} \cdot \nabla) \vec{u},
\label{eq4}
\end{equation}
\noindent where the r.h.s. of Eq.(3) is
nonzero only for an expansion/compression perpendicular to the direction of
\vec{B} and Eq.(4) shows that the ratio $B / \rho$
changes only in an expansion/compression parallel to $\vec{B}$.

With these relations in mind we consider the circumstellar medium of a
single massive star before it is reached by the SNR shock. The stellar
wind produces an expanding hot, low density bubble. This dynamical
evolution compresses the ambient interstellar medium (ISM) in an outer
shock wave in an approximately spherically symmetric manner. It is
clear that the compression of this external medium will create a
shocked shell with the gas density increasing radially outwards and,
over most of the solid angle (where the interstellar B-field is not
radial), also an increasing magnetic field strength. All pre-existing
density inhomogeneities in the external medium in the form of clouds
will by themselves have generally higher B-fields, wherever the
density is high. The subsequent compression by the wind bubble will
tend to enhance these correlated density and B-field
contrasts. Azimuthal density variations will probably form upon this
compression as well, by azimuthal instabilities of the radiatively
cooling swept-up shell, regardless of pre-existing clouds.  All this
implies spatially correlated variations of $\rho$ and $B$ in the
medium upstream of the SNR shock.

As the theoretical model of BV06 suggests, the remnant of the
subsequent SN explosion should be old enough that the swept-up mass of
interstellar matter dominates the mass enclosed in the wind bubble as
well as the ejected mass from the explosion. Then we have the
 situation that most of the emitting medium is a shock-modified ISM
that originally had a structure as described above. Its nonthermal
emission dominates the \gr emission as well as the synchrotron
emission of the SNR, as we observe it today.

\subsection{Injection and acceleration of nuclear particles at the SNR shock}
If the magnetic field lines in front of the accelerating forward SNR blast wave
are quasi-perpendicular to the shock normal, suprathermal ions cannot escape
from the downstream region (i.e. from the remnant interior) into the upstream
region to start being diffusively accelerated. On the other hand, nuclear
particles can be readily injected in quasi-parallel shock regions where the
field lines are more or less parallel to the shock normal. And the number of
ions injected into the acceleration from each gas volume is to a first
approximation proportional to the thermal gas density. In such field regions a
high energy density of accelerated nuclear particles will be built up by the
acceleration process. This particle population will try to escape into the
upstream medium and thereby excite streaming instabilities which lead to
magnetic field amplification in the associated magnetic flux tubes. The
enhanced field amounts at the same time to an increased target strength for
synchrotron emission by electrons, while the enhanced energetic nuclear
particle intensity will increase the number of inelastic collisions per unit
volume at given gas density. As already mentioned in the Introduction an
  inhibition of acceleration in quasi-perpendicular shock regions may not occur
  in the majority of quasi-perpendicular regions of the shock because of
  cross-field diffusion. However, in principle this effect is there and may
  occur especially at pre-existing interstellar clouds encountered by the SNR
  shock.

The injection of energetic electrons into the acceleration process is not very
well understood, and is possibly not correlated with the ion
injection. Electron injection might actually occur everywhere over the SNR
shock surface. It may or may not increase with the plasma density. 

Let us for a moment assume that electron injection
increases strictly monotonically with the upstream gas density.
Then we would expect a {\it nonlinear} correlation between the hard X-ray
  and the \gr emissions in a hadronic model (see also section 2.3), since
  electrons do not play a role in the acceleration energetics
  and the resulting shock modification, acting de facto as test
  particles. Therefore we expect the {\it amplitude} of the accelerated
  electron energy distribution to increase with $\rho$, even though its {\it
  form} is entirely determined by the accelerating ion component and is in fact
  equal to the {\it form} of the ion energy distribution.  The reason is that
  only the nuclear component -- with its dominant energy density -- determines
  the shock structure, and therefore the form of the accelerated energy
  spectrum of all ultrarelativistic particles whose energy is high compared to
  the proton rest energy, including the electrons (as long as their radiation
  loss can be neglected). In fact, the lack of saturation of the electron
  acceleration will allow the energy density of nonthermal energetic electrons
  to increase strictly monotonically with their injection rate, i.e. with gas
  density, in contrast to the saturation behavior of the nonthermal nuclear
  particles which limits their injection rate.

  Even if the electrons were injected -- and thus also accelerated --
  everywhere with the same intensity, then the amplification of the magnetic
  field by the ions in the quasi-parallel shock regions will automatically lead
  to an enhancement of the synchrotron emission in these regions, suggesting a
  correlation of the synchrotron emission with the hadronic \gr emission. This
  should be qualitatively similar to the case in the bipolar remnant of SN
  1006, even though probably less pronounced in wind bubbles. The bipolarity of
  SN 1006 has been discussed in detail by \citet{vbk03}. For an experimental
  discussion of the SN 1006 magnetic field configuration, see \citet{rbd04}.

Summarizing the arguments so far, we might expect a good and perhaps even
nonlinear spatial correlation between the synchrotron emission and the
production of energetic nuclear particles, as a result of pre-existing
correlated density and magnetic field strength inhomogeneities, as well as from
field amplification in quasi-parallel shock regions. However, this last
condition on nuclear particle injection implies that not all gas density
enhancements need to show up as regions of enhanced nuclear energetic particle
density and amplified magnetic field strength \citep{Plaga}. In other words,
not all clouds in the environment of \rxj need to be regions of enhanced
synchrotron and hadronic \gr emission. Some might just be shielded against the
injection of ions by an unfavorable magnetic field direction relative to the
shock normal.

\subsection{Hadronic \gr emission in the Quasi-Sedov phase}

The previous arguments did not consider the explicit modulation effect of
upstream density inhomogeneities in the upstream medium on the strength of
nuclear particle acceleration and on the resulting hadronic \gr emission. To
address this question we again make use of the assumption that the evolutionary
phase of \rxj is dominated by the swept-up mass from the wind bubble-structured
interstellar medium. Then the SNR interior should be subsonic with a roughly
{\it uniform} total gas and particle pressure $P_{tot} \propto \rho_1
V_\mathrm{s}^2$, as known from the Sedov solution in the case without CR
acceleration. Let us call this phase the Quasi-Sedov phase. If for example, in
a situation that deviates from spherical symmetry, the shock reaches regions of
different gas density over its surface, then the local shock velocity
$V_\mathrm{s}$ there will vary with the local upstream gas density
$\rho_1$ as $V_\mathrm{s}\propto \rho_1^{-0.5}$. In the Quasi-Sedov phase the
nuclear energetic particle pressure $P_\mathrm{c}$ is an only slowly varying
function of time, locally equal to several 10 percent of
$P_{tot}$ if ion injection is efficient 
\citep[e.g.][]{kbv05}. And this nonthermal energy density is
primarily concentrated in the highest-energy particles of the spectrum, i.e. in
the VHE range. This implies an essentially uniform VHE hadronic \gr emissivity
$q_{\gamma} \propto P_\mathrm{c}$ in these regions and therefore a local
hadronic \gr production rate $Q_{\gamma}= q_{\gamma} \rho_\mathrm{d} \propto
\rho_1$, since the downstream gas density $\rho_\mathrm{d}$ is a fixed fraction
of the upstream density $\rho_1$ for a strong shock.

Therefore the hadronic \gr emission is approximately proportional to the local
gas density.  Unless the electron acceleration is anticorrelated with the
upstream gas density, the X-ray synchrotron emission will be correlated with
the hadronic \gr emission, because $B$ and $\rho$ are almost always spatially
correlated.

\subsection{Spatial correlations in the leptonic scenario}
In the highly likely case that the leptonic \gr emission is dominated by the IC
emission, the \gr morphology is basically determined by the spatial
distribution of the radiating electrons. On the other hand,
the synchrotron emission is proportional to the product of the energetic
electron density and the local magnetic field energy. In the hypothetical case
that the hadronic \gr emission is negligible, the observations would of course
require a good spatial correlation between X-ray and IC \gr emission. And, the
better this correlation in such a leptonic scenario, the more the magnetic
field strength must therefore be spatially uniform.

This is a most unlikely situation for \rxj\, given the fact that the
gas density appears quite variable around the SNR shell and the
remnant appears to be interacting with at least some of these
molecular clouds, as CO-observations suggest \citep{Fukui,mor05}. This
makes a purely leptonic scenario already empirically highly unlikely,
independently of any theoretical arguments.

\section{Comparison of the latest \gr spectrum with the previous theoretical
  model of BV06}

The latest version of the H.E.S.S. \gr spectrum \citep{aha07a} 
presented in Fig.1 has not only increased statistical accuracy, but
also a flat (hard) spectral shape at the lowest energies of $~
250$~GeV, and a smooth extension and fall-off towards the 100 TeV
region. For the evaluation of this spectrum the reflectivity changes
of the H.E.S.S. mirrors have been taken into account, leading to a
roughly 15 percent increase in flux compared to the 2005
spectrum. 

Note that the theoretical \gr spectra, calculated in BV06 and also presented in
Fig.2, correspond to the assumptions that \rxj was a core collapse supernova SN
of type II/Ib with a massive progenitor and explosion energy
$E_{sn}=1.8\times10^{51}$~erg, that it has an age of $\approx 1600$~yr and is
located at a distance of $\approx 1$~kpc. Although this general scenario
corresponds to the conclusions of most other authors, significantly larger
distances also have been considered in the literature \citep{sla99}. For
details, see the discussion in BV06.

Since the theoretical model cannot well determine the spectral amplitude, for
the reasons given in BV06, the above-mentioned flux increase is not relevant in
a comparison of theoretical and observational spectra\footnote{This is
approximately also true if cross-field diffusion of the highest-energy nuclear
particles modifies the shock almost everywhere, as discussed in the
Introduction.}.

However, the {\it forms} of the spectra agree remarkably well. We note that the
inferred leptonic IC and Nonthermal Bremsstrahlung spectra are depressed by
more than an order of magnitude relative to the observed spectrum. They cannot
explain the observations, if the magnetic field is indeed amplified to the
degree assumed in the theory and supported by the upper limit for the thickness
of the synchrotron filaments which one can derive (BV06) from the
XMM-observations of this remnant by \citet{Hiraga}. The resulting lower limit
for the magnetic field strength of $65~\mu$G has recently been supported by
Chandra observation of fast variations of the hard X-ray emission in some spots
in the remnant, possibly showing the localized existence of even mG magnetic
field strengths \citep{uat08} (see, however also \citet{butt08}).

%-----------------------------------------------------------------------fig.1
\begin{figure*}
\centering
\includegraphics[width=0.9\textwidth]{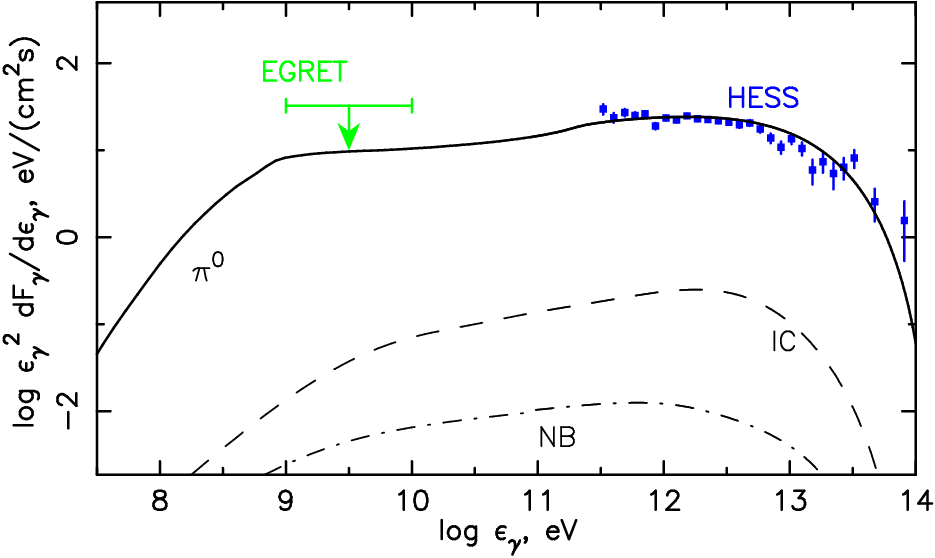}
\caption{Spatially integrated, \gr spectral energy distribution of \rxj . The
latest \hess~\gr data
\citep{aha07a} ({\it blue color}) are shown together with the EGRET upper limit
for the \rxj position \citep{aha06} ({\it green color}). The {\it
solid} curve at energies above $10^7$~eV corresponds to $\pi^0$-decay \gr
emission, whereas the {\it dashed} and {\it dash-dotted} curves indicate the
Inverse Compton (IC) and Nonthermal Bremsstrahlung (NB) emissions,
respectively, from the theoretical model of BV06. See also \citet{bv07} .}
\label{f1}
%\label{latest_\gr_spectrum_of_\rxj}
\end{figure*}
%------------------------------------------------------------------------------

\section{Comparison of the latest overall nonthermal spectrum with the BV06 
spectrum}

We present in Fig.2 along with new HESS data the latest hard X-ray spectrum,
obtained with the Suzaku instrument \citep{ttu08}, which is given in the form
of an energy flux density by \citet{uat08}, and compare it with the theoretical
spectrum of BV06 \citep[see also][]{za07}. The Suzaku measurement does not
cover the entire SNR, and therefore it needs to be suitably normalized by the
requirement of optimum agreement with the ASCA spectrum, cf. \citet{aha06},
over the latter instrument's range between 0.5 and 10 keV. The result is shown
in Fig.2. The good agreement basically stems from the fact that the amplified
downstream field of $\approx 100~\mu$G, used to calculate the theoretical
spectrum in BV06, already leads to electron synchrotron cooling above an
electron momentum of $\approx 500~m_\mathrm{p} c$, and thus to a {\it
hardening} of the synchrotron spectrum compared to a spectrum calculated
without electron cooling (see Fig.3 below).

%-----------------------------------------------------------------------fig.2
\begin{figure*}
\centering
\includegraphics[width=0.9\textwidth]{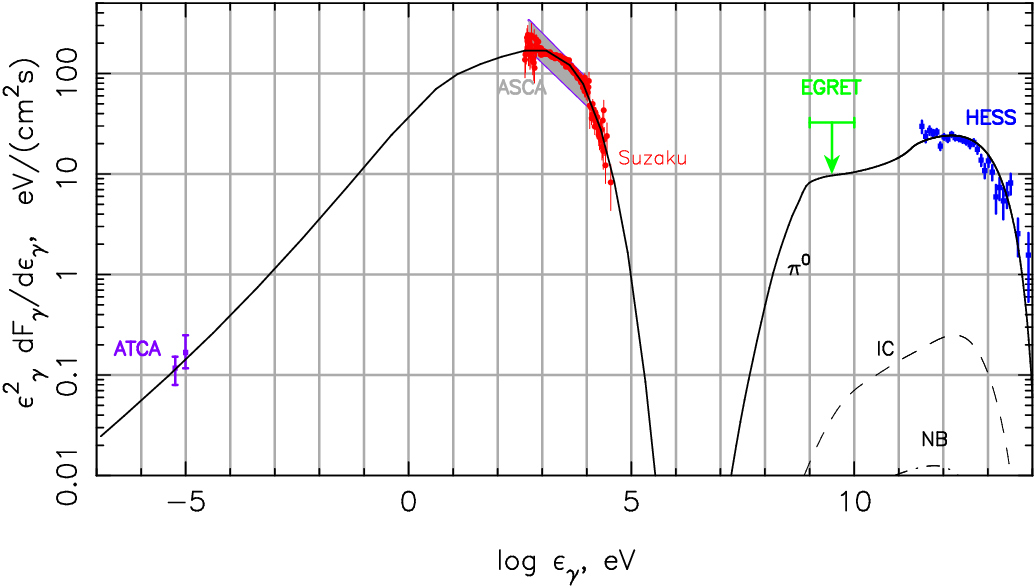}
\caption{Spatially integrated, overall nonthermal spectral energy distribution
  of \rxj\. The ATCA radio data \citep[cf.][]{aha06}[violet color], ASCA X-ray
  data \citep[cf.][]{aha06}, Suzaku X-ray data \citep{uat08}[red color], and
  2006 \hess \gr data \citep{aha07a}[blue color] are shown. The EGRET upper
  limit for the \rxj position \citep{aha06} [green color] is shown as well. The
  {\it solid} curve at energies above $10^7$~eV corresponds to $\pi^0$-decay
  \gr emission, whereas the {\it dashed} and {\it dash-dotted} curves indicate
  the Inverse Compton (IC) and Nonthermal Bremsstrahlung (NB) emissions,
  respectively, from the theoretical model of \citet{bv06}.}
\label{f2}
%\label{Overall_rxj_spectrum}
\end{figure*}
%------------------------------------------------------------------------------

The same observed spectrum can also be compared with a theoretical spectrum
(Fig.3) in which a very low ion injection rate ($\eta = 10^{-5}$) and a rather
low downstream magnetic field strength of $20~\mu$G was assumed (see
BV06). This corresponds to a dominantly leptonic \gr test particle spectrum
without field amplification\footnote{In fact, the strength of the downstream
magnetic field might be even smaller by a factor of two or more, wherever the
shock is not locally parallel to the external field. However, adopting such a
small field would imply that even the gross amplitude of the maximum of the
observed \gr energy flux could not be fitted to the observations.}. The
IC-scattered diffuse radiation field is the CMB plus interstellar far infrared
and optical radiation fields as given in \citet{bpv03}. This corresponds to
typical values used for the Solar neighborhood
\citep[e.g.][]{drury94,gps98,Porter06}. We note that for nearby objects at
distances $d\sim 1$~kpc the CMB contribution dominates in the IC emission
\citep[see also][]{Porter06}.

The electron injection strength was fitted such that an optimum fit to the
observations in the radio and X-ray ranges is achieved, cf. Fig.3. We note that
the {\it form} of the X-ray spectrum is only very poorly fitted in this
leptonic scenario, especially when one uses the recent Suzaku
measurements. Also the \gr spectrum has a maximum which is much too sharp in
comparison with the observed H.E.S.S. spectrum. We note that, compared to
earlier measurements \citep{aha05}, the latest version of the H.E.S.S.  \gr
spectrum \citep{aha07a} deviates more clearly from the IC-type spectrum with a
relatively sharp peak at $\epsilon_{\gamma}\sim 1$~TeV. Note also that our
spectrum of the nonthermal emission, that corresponds to the leptonic (or
inefficient) scenario, is almost identical to the spectrum presented by
\citet{Porter06}, even though they approximate the electron spectrum
analytically, whereas we calculate it numerically. Therefore it is also clear
that the quality of the fit achieved by \citet{Porter06} with a leptonic model
will be considerably lower if one uses the Suzaku X-ray spectrum and the new
HESS \gr spectrum instead of older, less accurate data.

At \gr energies of 1 GeV the spectral energy flux density is a factor of about
30 below the value in the hadronic scenario. It might be possible to construct
a more or less plausible form of the diffuse radiation field spectrum to obtain
a reasonable fit in the TeV region. However, it remains very doubtful in our
view whether this can avoid the sharp decline towards the GeV energy range
indicated in Fig.3. In any case such a construction cannot improve the
unacceptable fit in the hard X-ray range.

%-----------------------------------------------------------------------fig.3
\begin{figure*}
\centering
\includegraphics[width=0.9\textwidth]{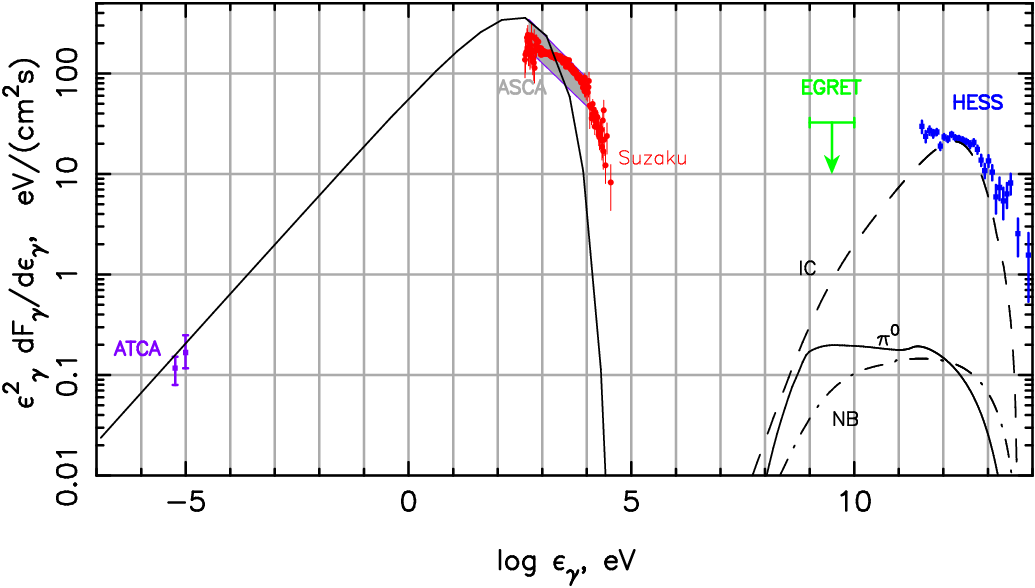}
\caption{The same as in Fig.2, except that a leptonically dominated scenario
was assumed (see text).}
\label{f3}
%\label{Inefficient_overall_rxj_spectrum}
\end{figure*}
%------------------------------------------------------------------------------

\section{Conclusions}
We conclude that a theoretical acceleration model which takes into account
magnetic field amplification and a consistent nuclear energetic particle
production is consistent with the latest H.E.S.S.  \gr and Suzaku hard X-ray
observations. It is also expected to be consistent with the observed good
correlation between X-ray synchrotron emission and VHE \gr emission. In the
face of existing gas density variations in or near the SNR, a purely leptonic
interpretation of the spatial correlation becomes poorer as this correlation
becomes closer empirically. The attempt to explain the \gr emission by leptonic
processes cannot be made consistent with the observed synchrotron spectrum. The
availability of higher-quality hard X-ray and \gr measurements leads to a
better consistency with the kinetic nonlinear theory prediction, whereas the
consistency of the leptonic model becomes much poorer.

Whether a leptonic model can lead to an acceptable fit of the \gr spectrum
through a more detailed evaluation of the local diffuse radiation field is an
open question. In our view, even such a modification would encounter enormous
difficulties in attempting to fit the \gr spectrum over the additional two
orders in \gr energy down from the VHE range to the GeV range. The leptonic
scenario seems also to be inconsistent with the filamentary X-ray morphology
which suggests substantial field amplification at least over part of the
remnant.

\begin{acknowledgements}
The authors would like to thank V.S. Ptuskin, F.A. Aharonian and
V.N. Zirakashvili for discussions on the general topic. EGB
acknowledges the partial support by the Presidium of RAS (program
No.16) and by the Russian Foundation for Basic Research (grant
07-02-00221) and the hospitality of the Max-Planck-Institut f\"ur
Kernphysik, where part of this work was carried out.
\end{acknowledgements}

\end{document}